\documentclass[aps,prb,reprint,superscriptaddress]{revtex4-1}
\usepackage{graphicx,amssymb,amsmath,amsfonts,verbatim,color,xr}

\bibpunct{[}{]}{,}{n}{}{}

\begin{document}
\bibliographystyle{apsrev}

\title{Vortex crossing, trapping and pinning in superconducting nanowires of a NbSe$_2$ two-dimensional crystal}

\author{Shaun A. Mills}
\author{Jacob J. Wisser}
\affiliation{Department of Physics and Materials Research Institute, The Pennsylvania State University, University Park, PA 16802, USA}

\author{Chenyi Shen}
\author{Zhuan Xu}

\affiliation{Department of Physics, Zhejiang University, Hangzhou 310027, China}
\affiliation{Collaborative Innovation Center of Advanced Microstructures, Nanjing 210093, China}

\author{Ying Liu}
\email{yxl15@psu.edu}

\affiliation{Department of Physics and Materials Research Institute, The Pennsylvania State University, University Park, PA 16802, USA}

\affiliation{Collaborative Innovation Center of Advanced Microstructures, Nanjing 210093, China}

\affiliation{Department of Physics and Astronomy and Key Laboratory of Artificial Structures and Quantum Control 
(Ministry of Education), Shanghai Jiao Tong University, Shanghai 200240, China}

\begin{abstract}
Nanowires of two-dimensional (2D) crystals of type-II superconductor NbSe$_2$ prepared by electron-beam lithography were studied, focusing on the effect of the motion of Abrikosov vortices. We present magnetoresistance measurements on these nanowires and show features related to vortex crossing, trapping, and pinning. The vortex crossing rate was found to vary non-monotonically with the applied field, which results in non-monotonic magnetoresistance variations in agreement with theoretical calculations in the London approximation. Above the lower critical field, $H_{c1}$, the crossing rate is also influenced by vortices trapped by sample boundaries or pinning centers, leading to sample-specific magnetoresistance patterns. We show that the local pinning potential can be modified by intentionally introducing surface adsorbates, making the magnetoresistance pattern a ``magneto fingerprint'' of the sample-specific configuration of vortex pinning centers in a 2D crystal superconducting nanowire.
\end{abstract}

\maketitle

\section{Introduction}
\label{intro}

The dynamic behavior of Abrikosov vortices has long been a subject of fundamental interest. The topic gained increased attention due to its relevance to practical applications of high temperature superconductors\cite{BlatterRMPVortices}. Recently, there has been interest in vortex dynamics in nanoscale systems, in particular, in manipulating individual vortices, which may enable experiments quantifying fundamental properties of Abrikosov vortices such as the vortex mass\cite{SuhlVortexMass,ChudnovskyVortexMass,FilVortexMass,GolubchikVortexMass} and forces influencing vortex motion\cite{HuebenerBook} (e.g., damping, pinning, boundary, and Magnus). Novel superconducting devices exploiting both the classical\cite{HastingsVortexRatchet,BerciuVortexSpintronics} and quantum\cite{GrosfeldAC, AoVortexInterference} motion of individual Abrikosov vortices were proposed.

The manipulation of individual vortices within doubly-connected superconducting nanoloops was reported recently\cite{MillsNbSe2Loop}. In that system, vortex motion is detected through its influence on the periodic magnetoresistance oscillations. In this Article, we explore the influence of vortices on the magnetoresistance of singly-connected nanowires. In the one-dimensional limit, the resistance of a superconductor is controlled by the rate of phase slips -- $2\pi$ jumps of the phase of the superconducting order parameter\cite{LAMH1,LAMH2}. The rate of phase slip events is determined by either thermal activation over\cite{TinkhamSnPhaseSlips,RogachevMagFields, RogachevNbNW} or, in the low-temperature limit, quantum tunneling through\cite{GiordanoMQT,BezryadinMoGeMQT,LauMQT,AltomareMQT} an energy barrier, leading to an exponentially small resistance below the critical temperature, $T_c$. In 2D narrow strips, instead of the magnitude of the order parameter fluctuating to zero at a single point and allowing for a phase ``unwinding,'' an Abrikosov vortex can nucleate at a sample boundary, cross the strip, and exit the opposite side\cite{Qiu2DPhaseSlips,BerdiyorovTDGL,BerdiyorovNbLadder}, carrying with it a $2\pi$ phase slip. The voltage induced by phase slips is given by the Josephson relation, $\partial\phi/\partial t=2eV/\hbar$\cite{Tinkham}. As in the 1D case, the phase slip rate, and therefore the induced voltage drop, is also determined by a corresponding energy barrier.

For a spatially isolated single vortex crossing event, this energy barrier can be calculated in the London approximation\cite{KoganLondon}. A major advantage of the London approximation is its validity over a wide range of temperatures, unlike the Ginzburg-Landau theory, which is formally restricted to temperatures very near $T_c$. Anticipating that potential quantum applications of vortex manipulation will necessarily be performed at temperatures far below $T_c$, we chose to investigate the magnetoresistance signatures of vortex dynamics in an experimental system far from $T_c$; we therefore interpret our observations in accordance with the London formalism. As we show below, in certain field ranges, the energy barrier determining the vortex crossing rate varies non-monotonically, giving rise to non-monotonic magnetoresistance variations in the nanowire. These magnetoresistance variations are distinct from the critical current oscillations observed in nanowires of disordered aluminum\cite{Morgan-WallWebberBlockade} and \emph{a}:InO\cite{JohanssonNWSQUID}, which are due to the formation of a ``Webber blockade''\cite{PekkerBlockade,AtzmonBlockade}, wherein static vortices are added to a nanowire one at a time in analogy to the Coulomb blockade in quantum dots\cite{FultonQuantumDot}.

The London approximation has some established limitations; most notably, it neglects the energy contributions arising from the finite extent of the vortex core and is restricted to systems without large gradients in field or Cooper pair density. In systems with a short coherence length, $\xi$, and large penetration depth, $\lambda$, (i.e., strongly type-II systems), these limitations are overcome. To address the intermediate regime ($\lambda\sim5\xi$), previous authors have investigated the magnetoresistance of superconducting nanowires utilizing a time-dependent Ginzburg-Landau (TDGL) phenomenology\cite{BerdiyorovTDGL}. The Ginzburg-Landau approach allows for treatment of systems where spatial variations in the order parameter are significant, such as is the case with a high density of vortices. While the TDGL equations are only rigorously justified in the case of a gapless superconductor\cite{Tinkham}, the results obtained are often in qualitative agreement with experimental observations. For the 2D crystal nanowire discussed in this Article, we show below that the London approach predicts the same influence of vortex dynamics on the magnetoresistance as the TDGL approach -- namely, that crossing Abrikosov vortices generate a non-monotonic magnetoresistance signature in nanowires.

Below, we present magnetoresistance measurements on long superconducting nanowires of 2D crystal NbSe$_2$. The nanowires have a width of $\lesssim10\xi(0)$, which is narrow enough that the vortex crossing mechanism is expected to dominate the magnetoresistance through the generation of phase slips\cite{Qiu2DPhaseSlips}, but wide enough to also allow for the trapping and pinning of vortices within the strip\cite{VodolazovStrip}. The nanowires have a length of $\gtrsim200\xi(0)$, placing them in a regime where spatial variations in the crossing barrier along the length of the nanowire affect the magnetoresistance signature. In this regime, we observe aperiodic mangetoresistance variations arising from vortex crossing events, in agreement with theoretical calculations in the London approximation (see Section~\ref{Theory}). Above a critical magnetic field, the presence of geometrically trapped and weakly pinned vortices locally modifies the barrier for vortex crossing and the resulting magnetoresistance variations. In this way, the magnetoresistance variations provide a ``magneto fingerprint'' that serves as a map of the sample-specific pinning potential. By intentionally modifying the pinning potential through the addition of surface adsorbates, we change the magnetoresistance response of the nanowire. 

\section{Vortex trapping, crossing, and pinning in nanowires}
\label{Theory}

The stability of a single Abrikosov vortex within a 2D nanowire has been considered theoretically on many occasions\cite{LikharevSCWire,KoganLondon,PekkerBlockade,AtzmonBlockade,VodolazovStrip}. The London equation governing the local magnetic field, $\vec b$, inside a thin film containing a single vortex at position $\vec v$ is given by
\begin{equation}
\vec{b}+\frac{4\pi\lambda^2}{c}\nabla\times\vec j=\phi_0\hat z\delta(\vec r-\vec v),
\label{London}
\end{equation}
where $\vec j$ is the current density and $\phi_0=h/2e$ is the magnetic flux quantum. We note that Eq.~\ref{London} fails within approximately $\xi$ of the vortex position, but for the experimental system we consider, $w\sim10\xi$, so the modification to the final result is minor\cite{Qiu2DPhaseSlips,SochnikovNNano}. Kogan, et al. provide an analytical solution for $\vec j$ in the case of a loop of inner radius $a$ and outer radius $b$\cite{KoganLondon}. By taking the limit as $a,b\to\infty$ while $b-a=w$ is held constant, we extend their result to the case of an infinitely long nanowire of width $w$. Once $\vec j$ is known, the free energy of the system can be calculated as the sum of the kinetic and magnetic contributions.

\begin{figure}
\begin{centering}
\includegraphics[width=\linewidth]{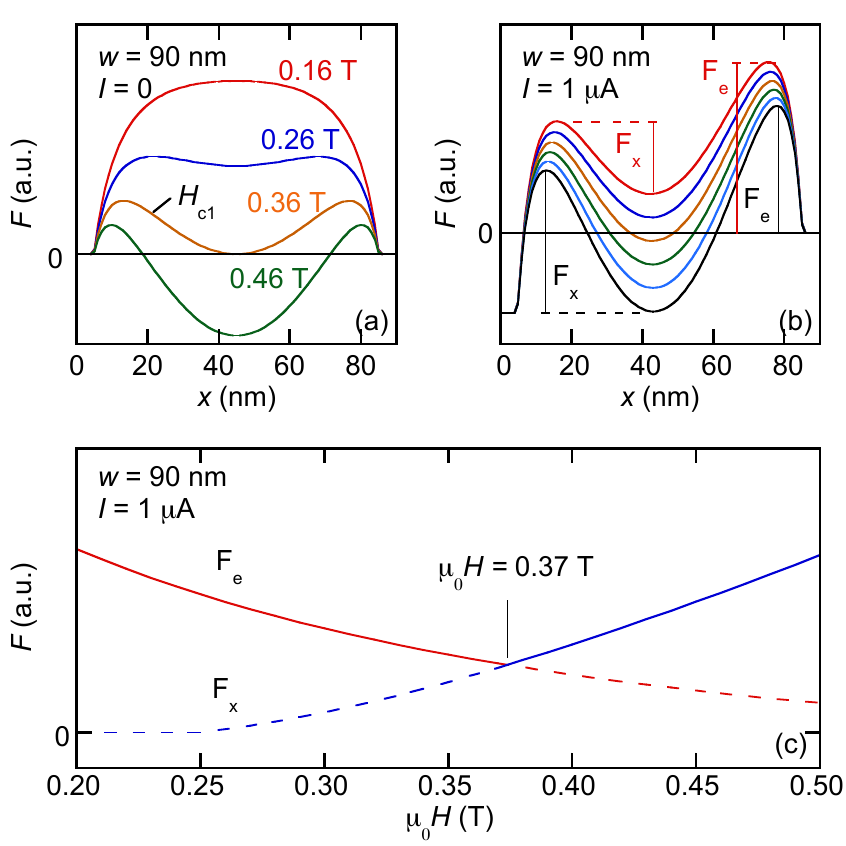}
\caption{(a) Vortex energy ($F$) vs. position along nanowire width ($x$) under applied fields of 0.16, 0.26, 0.36, and 0.46~T, top to bottom. Plot generated assuming $\xi=9.6$~nm, $\lambda=200$~nm, $d=9$~nm, $w=90$~nm, and $T=1.8$~K. (b) $F(x)$ for same nanowire in (a) at 1~$\mu$A applied current and applied fields of 0.33 to 0.38~T (top to bottom). Vortex flow is right to left. The barriers for vortex entry ($F_e$) and vortex exit ($F_x$) are indicated for 0.33 and 0.38~T. (c) $F_e$ and $F_x$ vs. applied magnetic field ($\mu_0H$) for nanowire in (a). At a given field, the larger of the two barriers (solid lines) dictates the vortex crossing rate.}
\label{wire-Hc1}
\end{centering}
\end{figure}

In Figure~\ref{wire-Hc1}(a) we plot the free energy, $F$, of a nanowire as a function of vortex position along the width of the wire at various applied magnetic fields in the absence of an applied current. The parameters used to generate these curves mimic the NbSe$_2$ nanowire we consider in Secs.~\ref{methods}-\ref{pinning}, namely, $\xi=9.6$~nm, $\lambda=200$~nm, $d=9$~nm, and $w=90$~nm, where $d$ is the nanowire thickness and $w$ is the nanowire width. In the case of zero applied current, the free energy acquires a global minimum within the nanowire at a critical field, $H_{c1}$. This is the field at which a vortex can first be trapped within the nanowire\cite{StanWireHc1}. Vortex trapping is a result of a global free energy minimum arising purely from the interplay between screening currents and the vortex self-currents, and is distinct from the concept of vortex pinning by defect sites discussed below. For our experimental system, we obtain a trapping field of $\mu_0H_{c1}=0.36$~T in the absence of an applied current (see Fig~\ref{wire-Hc1}(a)).

An external transport current will introduce an additional energy term to the curves in Fig.~\ref{wire-Hc1}(a), because a net current exerts a transverse Lorentz force on a vortex\cite{Tinkham}. If we assume any transport current is uniformly distributed within the nanowire, this results in the addition of a linear energy term, giving rise to the free energy shown in Figure~\ref{wire-Hc1}(b). In this case, the transport current flows into the page, applying a force in the negative $x$-direction on the vortex. The condition $F(x\leq0)<F(x>0)$ will cause a net flow of vortices across the width of the nanowire, provided the local energy barriers can be overcome. The rate of vortex flow is dependent upon the height of the two energy barriers shown in Fig.~\ref{wire-Hc1}(b); $F_e$ is the barrier for a vortex to enter the nanowire, and $F_x$ is the barrier for a vortex to exit the nanowire. Because a vortex must both enter and exit the nanowire to complete a cycle, and the probability of surmounting an energy barrier is exponentially dependent upon the barrier height, the crossing rate is primarily determined by the larger of the two barriers. In Figure~\ref{wire-Hc1}(c), we plot $F_e$ and $F_x$ for a nanowire with 1~$\mu$A applied current as a function of external field. The barrier which most influences the crossing rate at a given field is drawn with solid lines. At $\mu_0H=0.37$~T, the two barriers are equal, resulting in a local maximum in the crossing rate. It is this local maximum that leads to the non-monotonic magnetoresistance of the nanowire as discussed in Section~\ref{signatures}. 

It was previously demonstrated in 2D crystal nanoloops of NbSe$_2$ that vortex trapping and vortex crossing events can coexist under appropriate conditions\cite{MillsNbSe2Loop}. At higher fields (not shown), the curves in Fig.~\ref{wire-Hc1}(b) will again develop a global minimum within the nanowire. At this point, vortices can be trapped, and the barrier for additional crossing vortices will be modified by the interaction with the trapped vortices. The present theory is not suited to quantitatively address this situation. The curves in Fig.~\ref{wire-Hc1} assume a spatially isolated vortex, which is a reasonable assumption at low fields, where overlapping vortices are energetically unfavorable. However, it has been demonstrated in geometrically confined systems that giant multi-quanta vortices can form in certain field ranges\cite{ChibotaruAlSquare,CrenPbIslands2}. Thus, it is conceivable that a vortex may cross the nanowire along a path which passes near, or even intersects, a trapped or pinned vortex. We will return to a discussion of this situation in Sections~\ref{variations} and \ref{pinning}.

In addition to geometric constraints, which provide a means of vortex trapping, experimental 2D systems also feature finite levels of disorder, which result in vortices being preferentially pinned at locations where the order parameter is artificially reduced\cite{BlatterRMPVortices}. The strength of this pinning force decreases with decreasing crystal thickness for relatively thick NbSe$_2$ samples\cite{GoaNbSe2Pinning}, though there have been no systematic measurements of the pinning force in ultra-thin 2D crystals. Weak pinning centers need not necessarily immobilize a vortex, but they will locally lower the free energy of the vortex, which can result in either an effective increase or decrease in the crossing barrier depending on the location of the pinning center within the nanowire. 

\section{Magnetoresistance signatures of vortex dynamics}
\label{signatures}

The presence of crossing, trapped, and pinned vortices in a nanowire can be inferred from magnetoresistance measurements. Crossing vortices induce a transverse voltage drop due to the Josephson relation. For a time-averaged measurement, the voltage drop along the length of the nanowire will be proportional to the rate of vortex crossing. The non-monotonic dependence of the crossing rate on applied field arises from the non-monotonic crossing barrier height seen in Fig.~\ref{wire-Hc1}(c), and results in the magnetoresistance variations which are the focus of this Article.

The effect of static vortices on the magnetoresistance signatures is perhaps less intuitive, as static vortices produce no voltage signal on their own\cite{Tinkham}. However, the presence of static vortices has previously been shown to influence the motion of dynamic vortices, and in that way produce an effective magnetoresistance signature. In the context of nanoloops, dynamic vortices were shown to produce large amplitude periodic magnetoresistance oscillations\cite{SochnikovNNano,CaiTwoPeriods}, and the oscillations acquired a discrete phase shift at the field at which static vortices were trapped in the loop\cite{MillsNbSe2Loop}. Similarly, in the Webber blockade picture\cite{Morgan-WallWebberBlockade}, the periodic free energy modulation from the quantized addition of static vortices to the system results in periodic critical current oscillations\cite{PekkerBlockade,AtzmonBlockade}. In the context of the nanowires considered here, above $H_{c1}$, trapped and pinned vortices will clearly modify the local crossing barrier and in that way generate an effect on the magnetoresistance variations.

The curves shown in Fig.~\ref{wire-Hc1} represent the free energy of a single vortex within a nanowire. The vortex energy does not depend upon its position along the nanowire length, as the system is considered to be uniform in that direction. But experimental devices always feature inhomogeneities, so there will be some preferential location for vortex trapping due to local variations in the nanowire width and location of defect sites. Above $H_{c1}$, vortices will first become trapped at these preferential sites. Subsequent crossing vortices will interact with the trapped vortices as well as the screening and transport currents, and the free energy of the crossing vortices will not be adequately described by the curves in Fig.~\ref{wire-Hc1}. The addition of pinning sites featuring a suppressed order parameter further perturbs the energy calculation. However, the crossing rate will still be determined by the effective barriers, which will exhibit non-monotonic behavior due to a similar interplay between screening currents and vortex self-currents as before. These variations will be sample-specific, reflecting the sample-specific nature of the inhomegeneties. The preceding argument is phenomenological in nature, but as we show in Secs.~\ref{variations} and \ref{pinning}, it is substantiated by our experimental measurements and complementary to TDGL simulations.

\section{Experimental Methods}
\label{methods}

The nanowires presented in this article were fabricated from few-layer NbSe$_2$ crystals mechanically exfoliated from a bulk single crystal\cite{MillsNbSe2Fab}. Bulk NbSe$_2$ is a layered type-II superconductor with $\xi\approx10$~nm, $\lambda\approx200$~nm, and $T_c=7.1$~K\cite{SanchezNbSe2Quantities}. Each unit cell consists of two molecular layers in an AB stacking with an interlayer separation of 0.65~nm\cite{MarezioNbSe2Lattice}. Adjacent layers are weakly van der Waals coupled, allowing for easy cleaving with traditional mechanical exfoliation techniques\cite{NovoselovPNAS,StaleyNbSe2}. Bulk NbSe$_2$ features low intrinsic vortex pinning\cite{HessNbSe2}, but it is unclear what the pinning strength may be in the few-layer limit. One would expect the pinning force to decrease with decreasing sample thickness\cite{GoaNbSe2Pinning}, however, in the 2D limit, the effect of disorder may be enhanced.

\begin{figure}
\begin{centering}
\includegraphics[width=\linewidth]{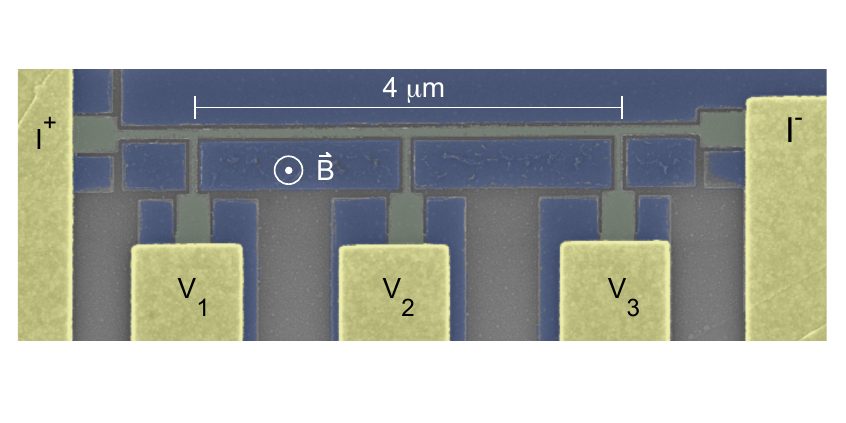}
\caption{(color online) False-color scanning electron micrograph of NbSe$_2$ nanowire. Functional device (light green) is isolated from surrounding NbSe$_2$ flake (dark blue) by 40~nm wide trenches (visible as dark outline). Yellow rectangular regions are Ti/Au electrical leads. Four-terminal measurement geometry is indicated.}
\label{SEM}
\end{centering}
\end{figure}

We first use optical microscopy to locate a suitable NbSe$_2$ flake on a Si/SiO$_2$ substrate. The height of the flake can be estimated to within one unit cell using a color code established by atomic force microscopy measurements\cite{MillsNbSe2Fab,StaleyNbSe2}. We then define measurement electrodes using electron beam lithography (EBL) and liftoff techniques. The electrodes are typically 30~nm gold with a 5~nm titanium adhesion layer. We outline our specific device with a second EBL write. By using a single layer of PMMA resist and a high resolution electron beam of $\sim7$~nm diameter (Vistec EBPG 5200 system), we can pattern feature sizes below 30~nm. Finally, we etch the excess flake with a CF$_4$ ion plasma to complete the fabrication of the nanostructure. The residual resist can be removed with a combination of a solvent bath and low-power O$_2$ plasma clean. Figure~\ref{SEM} shows a scanning electron micrograph of a five-terminal nanowire fabricated in this manner. The functional length of this device is 4~$\mu$m, the width is $90\pm4$~nm ($\sim9.5\xi$), and the thickness is $9\pm1.3$~nm. We fabricated three voltage leads to allow for independent measurements of the two segments of the full nanowire. 

All the measurements presented were performed in a Quantum Design Physical Property Measurement System with a base temperature of 1.8~K and a superconducting magnet capable of generating fields of $\pm9$~T perpendicular to the plane of the nanowire. Measurement leads were attached to the Ti/Au electrodes with a Kulicke \& Soffa Model 4123 ultrasonic wedge bonder, and extreme care was taken during wiring to prevent damage to the sample from electrostatic discharge. Transport measurements were carried out in a standard DC current-biased configuration using a Keithley 6221 current source and a Keithley 2182A voltmeter with measurement leads RF filtered at room temperature. Throughout this article, the term ``full nanowire'' will be used to indicate a measurement performed using voltage leads $V_1$ and $V_3$ (see Fig.~\ref{SEM}), whereas when the ``left'' (``right'') segment is discussed, it will denote a measurement performed using voltage leads $V_1$ and $V_2$ ($V_2$ and $V_3$).

\section{Magnetoresistance variations from vortex crossing}
\label{variations}

\begin{figure}
\begin{centering}
\includegraphics[width=\linewidth]{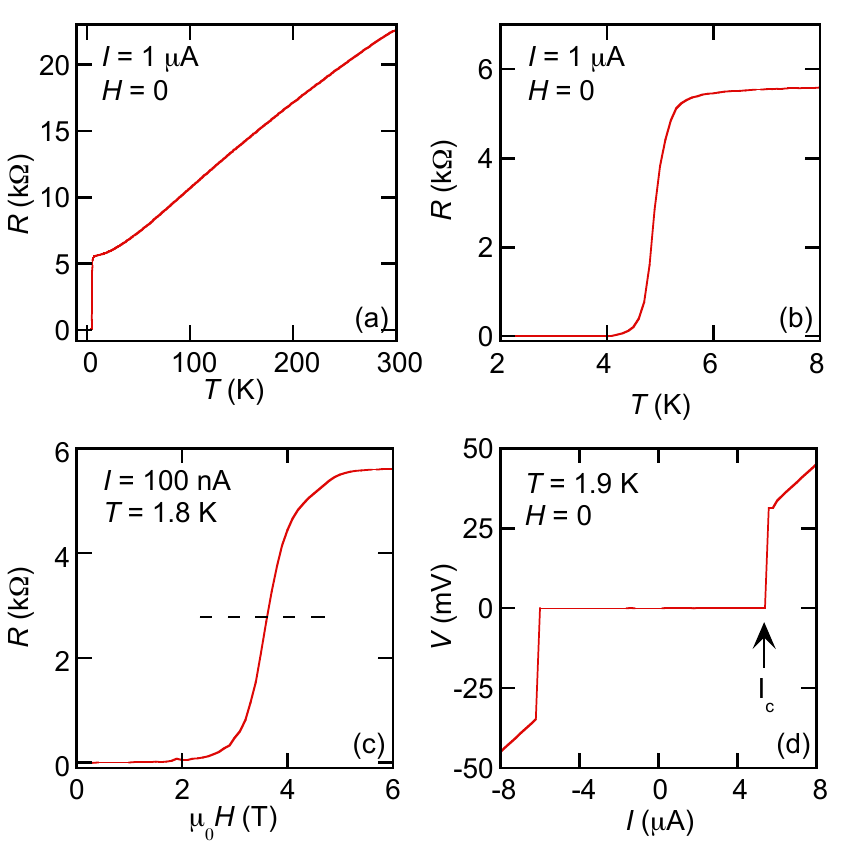}
\caption{(a) Resistance ($R$) \emph{vs.} temperature ($T$) of full nanowire over entire temperature range. (b) Low-temperature $R(T)$ at 0~T showing superconducting transition with onset transition temperature ($T_c$) of 5.6~K. (c) $R$ \emph{vs.} applied magnetic field ($\mu_0H$) at 1.8~K. Dashed line indicates $R=\frac{1}{2}R_N$, the criterion used to determine $\mu_0H_{c2}=3.6$~T. (d) Voltage ($V$) \emph{vs.} current ($I$) at 1.9~K and 0~T. The critical current, $I_c=5.5$~$\mu$A, is indicated.}
\label{characterization}
\end{centering}
\end{figure}

In Figure~\ref{characterization}, we present a general characterization of the full nanowire. The residual resistivity ratio, defined as $RRR=R(300K)/R(8K)$ is 4.0 for this device (Fig.~\ref{characterization}(a)), which is typical for this thickness of NbSe$_2$\cite{ElBanaFlakes}, and emphasizes the fact that the additional etch step does not substantially degrade the intrinsic quality of the single-crystal NbSe$_2$ flake. The device exhibits a superconducting transition with an onset $T_c$ of 5.6~K (Fig.~\ref{characterization}(b)). This is slightly reduced from the bulk value, but this reduction is again typical for thin flakes of NbSe$_2$\cite{StaleyNbSe2,FrindtNbSe2UnitCell,ElBanaFlakes}. In Figure~\ref{characterization}(c) we plot the resistance of the nanowire as a function of the perpendicular applied magnetic field (see Fig.~\ref{SEM} for field orientation) at 1.8~K. We define the critical field, $H_{c2}$, as the field at which the resistance is equal to one-half the normal state resistance. This criterion gives $\mu_0H_{c2}=3.6$~T, which, using standard Ginzburg-Landau relationships\cite{Tinkham} yields a coherence length of $\xi(1.8\;K)=9.6$~nm, consistent with the bulk value. We note that no magnetoresistance variations are visible in this full-scale measurement. Finally, in Figure~\ref{characterization}(d), we plot the current-voltage relationship for the nanowire in zero applied field. We observe a sharp transition to the zero-resistance state below an essentially symmetric critical current of $I_c=5.5$~$\mu$A. All subsequent measurements we present are performed at a measurement current $I_m\le1$~$\mu A$, significantly below $I_c$.

\begin{figure}
\begin{centering}
\includegraphics[width=\linewidth]{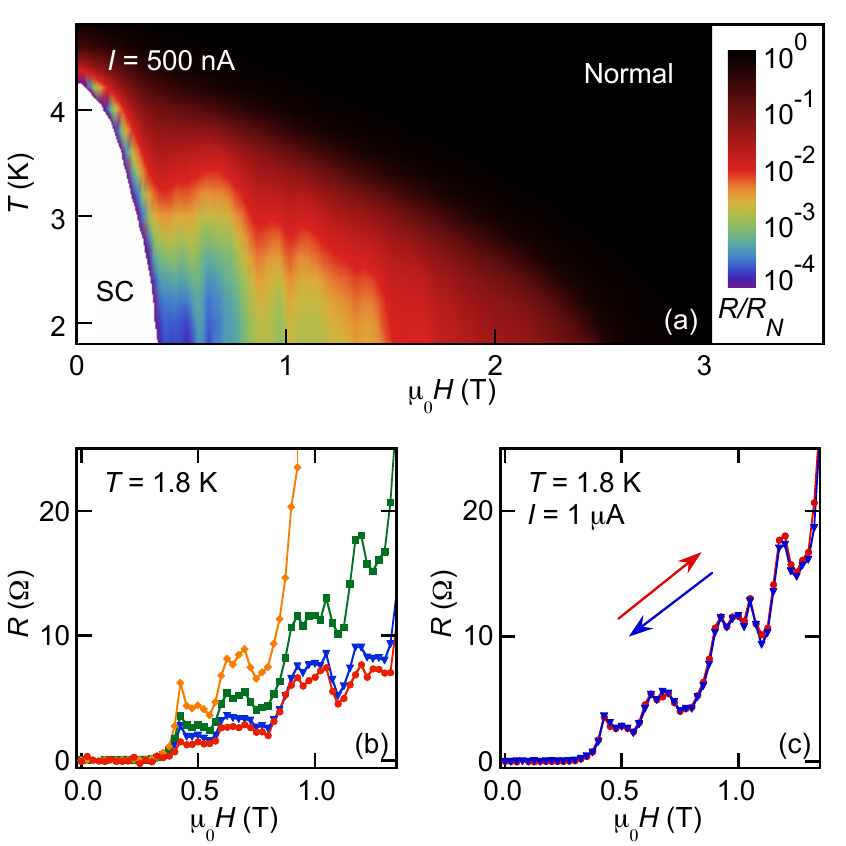}
\caption{(color online) (a) Color plot of resistance ($R$) \emph{vs.} magnetic field ($\mu_0H$) and temperature ($T$) at 500~nA. Resistance normalized to $R_N=5.5$~k$\Omega$ and plotted on log scale to emphasize variations. (b) $R(H)$ at 1.8~K and 0.1, 0.5, 1.0, and 1.5~$\mu$A (bottom to top). (c) $R(H)$ at 1.8~K and 1~$\mu$A in increasing (red circles) and decreasing (blue triangles) field.}
\label{MRV}
\end{centering}
\end{figure}

We now turn to low-field magnetoresistance measurements (see Fig.~\ref{MRV}). At a temperature of 1.8~K, the nanowire exhibits vanishing resistance in an applied magnetic field below 0.30~T. Above 0.30~T, we observe aperiodic magnetoresistance variations superimposed on an increasing background resistance. The locations of the relative extrema are insensitive to temperature (Fig.~\ref{MRV}(a)) and measurement current (Fig.~\ref{MRV}(b)) within our measurement resolution, and the first resistance maximum occurs at 0.42~T. These variations are not dependent upon field history (Fig.~\ref{MRV}(c)).

We first note that the magnetoresistance variations observed in our system are not consistent with the theory of a Webber blockade. In the Webber blockade picture, periodic magnetoresistance oscillations arise from the magnetic charging energy required to add an additional flux quantum to the nanowire, and therefore exhibit a period on the order of $\phi_0/A_0$, where $A_0$ is the area of the nanowire. For our system, that corresponds to a period of $\sim60$~Oe, which is more than an order of magnitude smaller than any reasonable definition of period from Figure~\ref{MRV}. Additionally, the variations remain quite pronounced even at temperatures above $0.5T_c$ (Fig. \ref{MRV}(a)), well beyond the point at which thermal fluctuations should mask any Webber blockade quantum oscillations\cite{JohanssonNWSQUID,PekkerBlockade}. This insensitivity to temperature and dramatically inconsistent ``period'' rule out the Webber blockade interpretation in this system.

On the other hand, the field-dependent behavior of the vortex crossing energy barrier shown in Fig.~\ref{wire-Hc1} explains the magnetoresistance variations both qualitatively and quantitatively. The observed resistance arises from crossing vortices, and is proportional to the crossing frequency, which depends upon the height of the two energy barriers. At 1.8~K,  the first resistance maximum is reached at 0.42~T, which is in agreement with the local minimum in the crossing barrier calculated to occur at $\mu_0H=0.37$~T in Fig.~\ref{wire-Hc1}(c). Subsequent resistance maxima at higher applied fields cannot be addressed quantitatively within the framework developed in Sec.~\ref{Theory}, which applies only to a spatially isolated crossing event. At higher fields, the vortex density will increase, and crossing events will no longer be spatially isolated, leading to a modification of the crossing barrier. In fact, at 0.37~T, simple geometric considerations suggest a chain of crossing vortices is located approximately every $6.5\xi$ along the length of the nanowire. It is reasonable to expect that the proximity of crossing events will modify the crossing barriers seen in Fig.~\ref{wire-Hc1}. As discussed in Sec.~\ref{signatures}, this should lead to additional magnetoresistance variations at higher fields.

The TDGL simulations of a superconducting nanowire reported previously by Berdiyorov et al.\cite{BerdiyorovTDGL} are relevant to the present study. The Ginzburg-Landau theory is not restricted to a regime of low vortex density. Berdiyorov observed that at low fields, vortices were excluded from the nanowire, whereas at high fields, vortices were immobilized within the nanowire. At intermediate fields, TDGL simulations revealed a series of chains of vortices periodically crossing the nanowire. This crossing cycle repeats with a frequency  $f^{-1}\sim100\tau_\text{GL}$, where $\tau_\text{GL}$ is the GL relaxation time, $\tau_\text{GL}=4\pi\lambda^2\sigma_n/c^2$ ($\sigma_n$ is the normal state conductivity). The system experienced two local maxima in $F(t)$ in the entry-crossing-exit process. These maxima were defined as effective entry and exit barriers. The frequency of vortex crossing was correlated with the relative heights of the two local maxima, and observed to vary non-monotonically with field in intermediate field ranges. At fields well above the first magnetoresistance maximum, TDGL simulations revealed additional magnetoresistance peaks. These subsequent peaks apparently resulted from additional chains of crossing vortices, and were typically superimposed upon a resistive background, indicating that at sufficiently high fields and current densities, vortices are always in motion, and not immobilized within the nanowire. Our observations in Fig.~\ref{MRV} are consistent with this prediction. The results of the TDGL simulations are complementary to the results in Sec.~\ref{Theory}. The London formalism describes two energy barriers in the space domain resulting from the interplay between applied, screening, and vortex self-currents, whereas the TDGL simulations observe two free energy maxima in the time domain resulting from a particular spatial distribution of the order parameter.

\section{Effect of pinning centers from surface adsorbates}
\label{pinning}

\begin{figure}
\begin{centering}
\includegraphics[width=\linewidth]{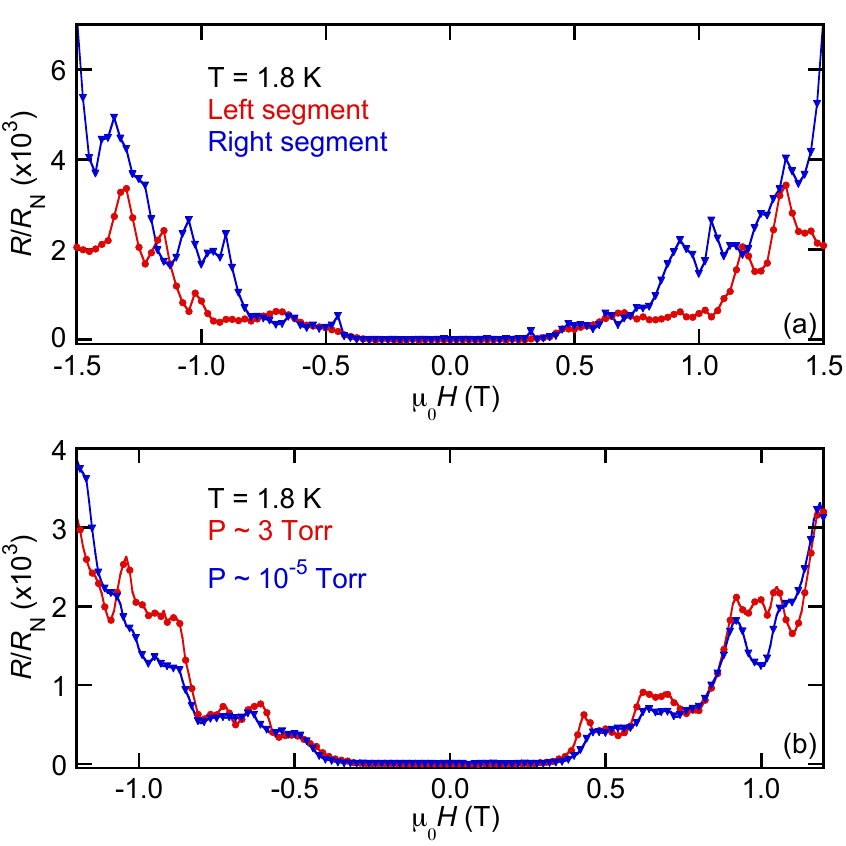}
\caption{(color online) (a) Normalized resistance ($R/R_N$) \emph{vs.} magnetic field ($H$) for two independent sections of the nanowire. Left segment (red circles) is measured from $V_1$ to $V_2$, and right segment (blue triangles) is measured from $V_2$ to $V_3$ (see Fig.~\ref{SEM}). (b) $R/R_N(H)$ for full nanowire at 1.8~K in gaseous $^4$He environment at pressures of 3~Torr (red circles) and $10^{-5}$~Torr (blue triangles).}
\label{differences}
\end{centering}
\end{figure}

It is clear that the idealized free energy shown in Fig.~\ref{wire-Hc1} will be modified by the presence of defect regions in experimental devices. We investigated this situation experimentally by comparing the magnetoresistance of two independent and nominally identical nanowires. In Fig.~\ref{differences}(a), we plot the magnetoresistance variations for the left and right segment of the nanowire shown in Fig.~\ref{SEM}. Both curves have been normalized to the normal state resistance of the respective segment, which differed by $<3$\%. Each segment exhibits magnetoresistance variations, and, significantly, the field at which the first resistance maximum appears is independent of nanowire segment and field sign. The constant value of this first maximum for the two segments is anticipated, as the effective width is essentially uniform along the nanowire, and in the London formalism, it is the nanowire width alone that determines the first crossing rate maximum. However, the amplitude and position of relative extrema appear uncorrelated between the segments. We note that a similar irreproducibility was observed in Pb nanowires\cite{WangPbBelts}, but their sample fabrication method did not allow measurements of independent nanowire segments as used in the present study.

Specifically, two nanowires made as identically as possible showed differing magnetoresistance patterns, which can only be ascribed to the influence of differing configurations of pinning sites on the free energy in Fig.~\ref{wire-Hc1}. This sample-specific response to magnetic field is also anticipated in the TDGL simulations. By including regions with a suppressed superconducting order parameter, Berdiyorov et al. were able to shift the location and amplitude of the original magnetoresistance peak, as well as introduce additional peaks at higher fields\cite{BerdiyorovTDGL}. The disordered region pinning sites were observed to form ``easy-flow'' channels where vortices preferentially crossed the nanowire. In the complementary language of the London formalism in Section~\ref{Theory}, the pinning sites modify the entry and exit barriers at certain locations along the nanowire length. The location of pinning sites varies between samples, which causes different configurations of ``easy-flow'' channels, and therefore sample-specific magnetoresistance variation signatures. This sample-specific pattern of the magnetoresistance variation, which should be seen only in sufficiently long nanowires where multiple independent vortex crossing locations can exist, can be called a ``magneto fingerprint'' of the sample-specific configuration of vortex pinning centers in a 2D crystal superconducting nanowire.

The ultra-thin nature of our device appears to make it sensitive to surface defects and adsorbates, consistent with the recent finding that surface contamination can suppress superconductivity in atomically-thin NbSe$_2$\cite{GeimGlovebox,MakCDW1LNbSe2}. We measured the full nanowire in two different ambient conditions -- a low vacuum helium environment ($P\sim3$~Torr) and a high vacuum ($P\lesssim10^{-5}$~Torr) environment. As shown in Figure~\ref{differences}(b), the magnetoresistance variations differ in the two situations. This suggests that surface adsorbates also affect the magnetoresistance variation.
This is consistent with the observed lack of hysteresis in the magnetoresistance variations seen Figure~\ref{MRV}(c), because we would expect surface defects to generate only a weak pinning potential in our sample, which is several unit cells thick.
It would be useful to correlate these measurements with a high-resolution magnetic imaging technique to determine precise vortex positions. Finally, we note that, while the magnetoresistance variations depend on the pinning potential, the initial appearance of finite resistance is dictated by $H_{c1}$, rather than by vortex depinning. This can be clearly seen in Fig.~\ref{MRV}(b), which shows the field at which finite resistance appears is essentially unaffected by the measurement current. 

\section{Conclusion}
\label{conclusion}

In summary, we observed magnetoresistance variations in 2D crystal superconducting nanowires of NbSe$_2$ arising from the flow of vortex chains across the nanowire. The vortex crossing rate varies non-monotonically with field in agreement with calculations of the crossing barrier performed in the London approximation. In the size regime we explored, the magnetoresistance variations are not periodic and are superimposed on an increasing background. The magnetoresistance variations first appear above a critical field for vortex trapping, $H_{c1}$, which is observed to be in agreement with theoretical predictions for a 2D nanowire. The magnetoresistance variations are further influenced by the presence of weak pinning sites caused by random surface adsorbates, which is consistent with time-dependent Ginzburg-Landau simulations. It is clear that magnetoresistance measurements can provide a means of mapping the sample-specific pinning potential in 2D crystal superconductors. 

\begin{acknowledgments}
The work at Penn State is supported by NSF under Grant number EFMA1433378 . Work in China was supported by MOST of China (Grant 2012CB927403) and NSFC (Grants 11274229 and 11474198). Nanofabrication was carried out at the Penn State MRI Nanofabrication Laboratory.
\end{acknowledgments}


\end{document}